%% file: IAU314_Alex_Binks.tex
\documentclass{iau}
\usepackage{graphicx}

\title[The Debris Disk Fraction for M-dwarfs in Nearby, Young, Moving Groups]
{The Debris Disk Fraction for M-dwarfs in Nearby, Young, Moving Groups}

\author[A. Binks]
{Alex Binks$^1$
}

\affiliation{$^1$Keele University\\email: {\tt a.s.binks@keele.ac.uk}\\
}

\pubyear{2015}
\volume{314}
\pagerange{119--126}
\jname{Young Stars \& Planets Near the Sun}
\editors{J. H. Kastner, B. Stelzer, \& S. A. Metchev, eds.}
\begin{document}

\maketitle

\begin{abstract}

I present the first substantial work to measure the fraction of debris disks for M-dwarfs in nearby moving groups (MGs). Utilising the $AllWISE$ IR catalog, 17 out of 151 MG members are found with an IR photometric excess indicative of disk structure. The M-dwarf debris disk fraction is $\lesssim 6$ per cent in MGs younger than 40\,Myr, and none are found in the groups older than 40\,Myr. Simulations show, however, that debris disks around M-dwarfs are not present above a $WISE$ $W1-W4$ colour of $\sim 2.5$, making calculating the absolute disk fractions difficult. The debris disk dissipation timescale appears to be faster than for higher-mass stars, and mechanisms such as enhanced stellar wind drag and/or photoevaporation could account for the more rapid decline of disks observed amongst M-dwarfs.


\end{abstract}

\firstsection 

\section{Introduction}

Observations of debris disks surrounding low-mass stars are rare. Whilst much work has focused on the disk frequencies for Solar-type stars, little is known of M-dwarf debris disk fractions, because most are too faint to be studied in open clusters, young field M-dwarfs are very rare and until recently few were confirmed as members of nearby, young moving groups (MGs). The $WISE$ infrared (IR) photometric catalogue has identified hundreds of IR excesses in nearby Solar-type objects (Patel, Metchev $\&$ Heinze 2014), however, very few are found in young, nearby M-dwarfs (Avenhaus, Schmid $\&$ Meyer 2012). Debris disk fractions for young M-dwarfs have been reported to be both smaller than FGK-types (Lestrade et al. 2009) and larger (Forbrich et al. 2008), however, these are limited by small number statistics. 

\begin{figure}[b]
\begin{center}
\includegraphics[width=5.4in]{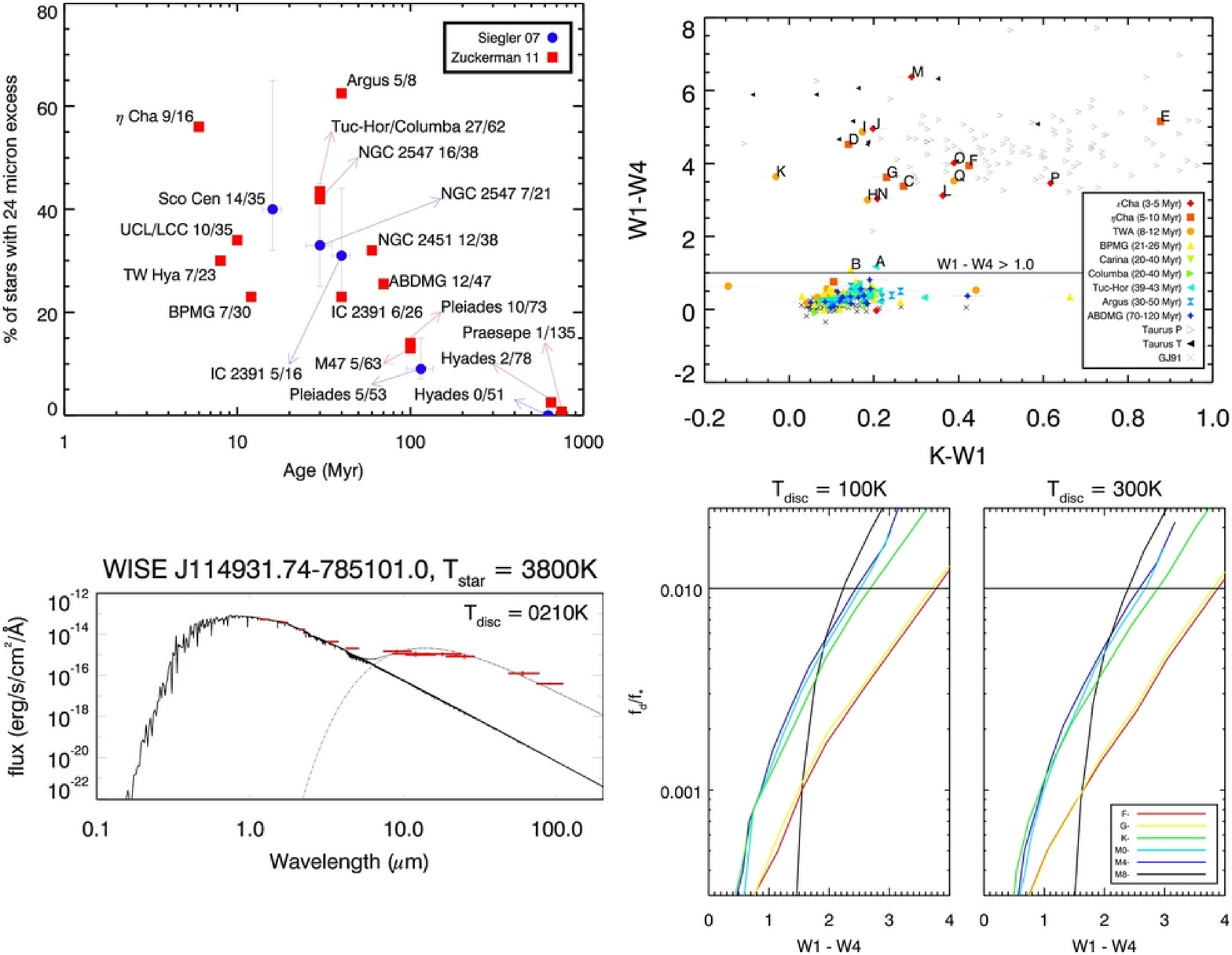} 
 \caption{Top left: the fraction of debris disks as a function of age in groups $\lesssim 650$\,Myr. Top right: $K-W1$ vs $W1-W4$ colour-colour diagram for all M-dwarfs in this sample. Bottom left: an example SED fit. Bottom right: $WISE$ sensitivity limits for debris disks.}
   \label{Disk_Fractions}
\end{center}
\end{figure}

The frequency of observed debris disks appears to be dependent on both age and spectral-type. Generally speaking, debris disks are more common around higher-mass, young stars (Wyatt 2008). The top-left panel in Fig.\,\ref{Disk_Fractions} shows that from 10 to 100\,Myr there is roughly an overall decline from $\sim 40$ to 10 per cent in FGK stars. Although some work has investigated the disk fractions in MGs (Simon et al. 2012; Schneider et al. 2012), few have been able to provide large M-dwarf samples. Motivated by the lack of a robust M-dwarf disk fraction and the recent identification of hundreds of new M-dwarf MG members (e.g., Malo et al. 2013; Gagn{\'e} et al. 2014), I use WISE photometry to attempt to measure the fraction of M-dwarfs with debris disks in MGs.

\section{Identifying and characterising disk excess}

All 151 candidates chosen for analysis have signal-to-noise ratios $> 5.0$ in the $WISE$ $W4$ band and have radial velocities to within $5\,{\rm km\,s}^{-1}$ of their proposed MG. The scatter in $W1-W4$ colour amongst M-dwarf field objects is $\sim 0.3$ therefore a cut of $W1-W4 > 1.0$ is chosen to select objects indicative of IR excess (c.f., Schneider et al. 2012), and the top right panel of Fig.\,\ref{Disk_Fractions} shows the 17 objects that qualify. Spectral class indices ($\alpha = d\log \lambda F_{\lambda}/d\log\lambda$) are calculated following the procedure in \cite{2012a_Riaz} and all objects with $W1-W4 > 1.0$ are either class II or III (see Table~\ref{Disk_Designation}).

Spectral energy diagram (SED) fits are made to stars with IR excess and the $IRAS$, $IRAC$, $MIPS$ and $AKARI$ catalogues were searched for additional data at longer wavelengths. Photospheric fits are generated from either the BT-Settl models ($T_{\rm eff} > 2700\,$K) or the AMES-DUSTY models (for $T_{\rm eff} \leq 2700\,$K) and single temperature black-body fits are made to represent a disk. The best-fit minimized $\chi^{2}$, however for 9 objects there are insufficient data at wavelengths $> 25\,\mu$m and the fit is degenerate. Disk flux fractions ($f_{\rm d}/f_{\rm bol}$) were calculated by dividing the disk flux by the stellar flux. Objects with $f_{\rm d}/f_{\rm bol} < 0.01$ are considered as debris disks (c.f., Lagrange et al. 2000), objects with $0.01 \leq f_{\rm d}/f_{\rm bol} < 0.1$ are classed as either primordial or transitional and $f_{\rm d}/f_{\rm bol} > 0.1$ is primordial. Column 6 in Table~\ref{Disk_Designation} provides the disk-type measured from SED fits. The bottom-left panel of Fig.\,\ref{Disk_Designation} shows an SED fit for an object with available IR data up to $100\,\mu$m. 

Each object was searched for any additional material in the literature that may point to the type of disk excess. Classification based on literary sources only assumes primacy if data has been used which wasn’t available for this work. Any parameters derived in this work are compared to results in the literature if they are available. Ultimately, the final designation of a disk type is subjective in some cases, particularly if there is a lack of mid-/far-IR data to constrain SED fits, however, if a sufficient amount of IR data is available then $f_{\rm d}/f_{\rm bol}$ is used as the primary indicator. In addition 7 objects are found with $W1-W4$ {\it less than} 1.0 but having literary sources claiming they are debris disks. SED fitting was appropriate for two of these objects, and both have $f_{\rm d}/f_{\rm bol} < 0.01$.
 
\input{Table1}

\section{Sensitivity limits using $WISE$}

Given that the objects in this work are relatively faint M-dwarfs, there will be a limiting $W1-W4$ for which objects can be observed with $f_{\rm d}/f_{\rm bol} \leq 0.01$. To explore the relationship between $f_{\rm d}/f_{\rm bol}$ and $W1-W4$, a simulation was made in which the emitting area of the disk was altered to give $f_{\rm d}/f_{\rm bol} = 0.01$, and this gave a corresponding $W1-W4$.

The results, displayed in the bottom-right panel of Fig.\,\ref{Disk_Designation}, show that the limiting $W1-W4$ colour corresponding to disk flux fractions $< 0.01$ is $\approx 2.5$ for early/mid M-dwarfs, $\approx 2.2$ for late M-dwarfs, and $\approx 3.5$ for Solar-type stars. However, if $W1-W4$ is too small it becomes indistinguishable from field stars. Therefore the simulations provide a `window' of $1.0 < W1-W4 < 2.5$ that can indicate debris disks between the error threshold at the bottom, to where $f_{\rm d}/f_{\rm bol} = 0.01$ at the top. In Table~\ref{Disk_Designation}, objects labelled with an asterisk in column 6 cannot be considered debris disks.

\section{Conclusions}

Only 2 of the objects with $W1-W4 > 1.0$ have $W1-W4$ values small enough to have $f_{\rm d}/f_{\rm bol} < 0.01$. Seven objects with $W1-W4 < 1.0$ were classed as debris disks based on literary sources and $f_{\rm d}/f_{\rm bol}$ (where possible). Only four MGs were found to have any debris disks; $\eta$ Cha, TWA, BPMG and Tuc-Hor. The total fraction of MG members with debris disks identified in this work is 9/151 (6 per cent), however, it can only be used as a lower limit as it is quite possible for debris disks to avoid detection with a very small $W1-W4$ excess. If the objects identified as transitional disks actually turned out to be debris disks, then these numbers change to 16/151 (11 per cent). Only five objects have a measurable disk flux fraction which is $< 0.01$. If only these objects were used, then a lower limit of 5/151 (3 per cent) would be set.

Within these narrow detection limits several stars were identified with a modest excess; too small to be a primordial disk, and in some cases there is longer wavelength data that supports the classification as a debris disk. Without knowing how many undetected debris disks have been unaccounted for it is difficult to provide a robust M-dwarf debris disk fraction. The fraction of debris disks is small, smaller than claimed for A--K stars at similar ages, but it is hard to compare the two because of a lack of comparable sensitivity to $f_{\rm d}/f_{\rm bol}$ for stars for different masses. Higher-mass stars contribute less relative flux at IR wavelengths and are more easily detectable from a field star population of similar spectral-type.

For objects younger than 40\,Myr, debris disks were detected in 9 out of 113 (8 per cent) and no debris disks were detected in the 38 objects older than 40\,Myr. This is a significant difference and the confidence to which one can claim a detection rate of $< 6$ per cent in M-dwarfs older than 40\,Myr is 90 per cent. Despite the sensitivity problems it is clear that the M-dwarf disks are evolving and that the evolutionary timescale for debris disk dissipation appears to be faster than for higher mass stars.

\begin{discussion}

\discuss{Lada}{I'd like to point out that we used {\it Spitzer} to observe some M-dwarf debris disks in NGC~2547 and actually the disk fractions turned out to be higher in M-dwarfs than for the Solar-type stars!}

\discuss{Author}{Yes, I would say that these results are not claiming debris disks don't exist but that they are not {\it observed}. Perhaps a deeper infrared survey of these objects would yield more disks.}

\discuss{Zuckerman}{Perhaps $WISE$ wasn't the ideal survey for you.}

\discuss{Author}{Maybe not, and like the simulations show, $WISE$ could only pick out disks in a very narrow colour band. But for now this is all we have to work on.}

\end{discussion}

\end{document}

%% file: Table1.tex
\begin{table}
  \begin{center}
  \caption{Final designation of disk-type based on photometric excess and SED modelling.}
  \label{Disk_Designation}
 {\scriptsize
  \begin{tabular}{lllllllr}\hline 
Name                  & Group          & W1$-$W4 & Excess & $\alpha$ & $f_{\rm d}/f_{\rm bol}$ & Literature & Final \\
\hline
J010335.75$-$551556.6 & Tuc-Hor        &    1.17 & TTT & II  & N               & N/A & D \\
J044356.87+372302.7   & BPMG           &    1.10 & TTT & II  & N/A             & N/A & D \\
J084130.24$-$785306.3 & $\eta$ Cha     &    3.39 & PPP & II  & 0.04, P/T       & P   & P \\
J084227.02$-$785747.7 & $\eta$ Cha     &    4.53 & TTT & III & 0.05, P/T       & P   & P \\
J084318.52$-$790518.0 & $\eta$ Cha     &    5.15 & PPP & II  & 0.54, P         & P   & P \\
J084409.09$-$783345.6 & $\eta$ Cha     &    3.94 & PPP & II  & 0.10, P/T       & P   & P \\
J084416.33$-$785907.8 & $\eta$ Cha     &    3.61 & TPP & II  & 0.03, P/T       & P   & P \\
J101209.04$-$312445.3 & TWA            &    3.01 & TPP & III & 0.02, P/T       & N/A & T \\
J111027.80$-$373152.0 & TWA            &    4.87 & TTT & II  & 0.09, P/T       & T   & T \\
J111835.64$-$793554.8 & $\epsilon$ Cha &    4.95 & TTT & III & N/A$^{*}$       & T   & T \\
J113218.24$-$301952.0 & TWA            &    3.64 & TPT & III & N/A$^{*}$       & D   & T \\
J114326.57$-$780445.5 & $\epsilon$ Cha &    3.12 & PPP & III & N/A$^{*}$       & T   & T \\
J114931.74$-$785101.0 & $\epsilon$ Cha &    6.38 & PPP & II  & 0.48, P         & P/D & P \\
J115504.71$-$791911.0 & $\epsilon$ Cha &    3.03 & TTP & III & N/A$^{*}$       & T   & T \\
J120055.08$-$782029.5 & $\epsilon$ Cha &    4.01 & PPP & II  & N/A$^{*}$       & P   & P \\
J120144.32$-$781926.7 & $\epsilon$ Cha &    3.46 & PPP & II  & N/A$^{*}$       & P   & P \\
J120733.42$-$393254.2 & TWA            &    3.53 & PPP & II  & N/A$^{*}$       & T   & T \\
\hline
\end{tabular}
  }
 \end{center}
\vspace{1mm}
 \scriptsize{
 {\it Notes:}
  D = debris disk, P = primordial disk, T = transitional disk, N = no evidence for a disk, N/A = no information available (*although there are no data available for SED fits beyond $25\,\mu$m the $W1-W4$ colour is beyond the limit for a debris disk and these can only be classed as either `P' or `T'. \\
}
\end{table}